\documentstyle[preprint,cite,epsf,aps,amssymb]{revtex}

\newcommand{\be}{\begin{equation}}
\newcommand{\ee}{\end{equation}}
\newcommand{\bea}{\begin{eqnarray}}
\newcommand{\eea}{\end{eqnarray}}
\newcommand{\ba}{\begin{array}}
\newcommand{\ea}{\end{array}}

\newcommand{\half}{\frac{1}{2}}

\begin{document}
\tightenlines

\title{On New Forms of the BRST Transformations}

\author{Victor O. Rivelles}

\address{Instituto de F\'\i sica, Universidade de S\~ao Paulo\\
 Caixa Postal 66318, 05315-970, S\~ao Paulo, SP, Brazil\\
E-mail: rivelles@fma.if.usp.br}

\maketitle

\begin{abstract}
We show how to derive systematically new forms of the BRST
transformations for a generic gauge fixed action. 
They arise after a symmetry of the gauge fixed action is found in the
sector involving the Lagrange multiplier and its canonical momentum. 
\end{abstract}

\vspace{1cm}

Every now and then new nilpotent symmetries are found for gauge
theories. Their transformation laws can be nonlocal and even not
manifestly covariant. In all cases it was possible to show that they
are different forms of the well known BRST symmetry
\cite{Rivelles:1995gb}. Usually, the claimed new transformations are
found in the Lagrangian formulation of the theory and it is quite
difficult to  
see its relation to the known form of the BRST transformations. Going
to the Hamiltonian formulation and using the BRST-BFV formalism
\cite{Fradkin:1975cq}, the origin of these different forms for the
BRST transformations can be fully understood. In
\cite{Rivelles:1995gb} it was also shown that there are symmetries of
the gauge fixed action involving only the ghost sector which gives
rise to these uncommon forms of the BRST transformations. In this
letter we will show that the sector consisting of 
the Lagrange multiplier and its momentum also has symmetries which
give rise to new forms for the BRST transformations.

Consider a gauge theory in the Hamiltonian formulation. After all
first class constraints have been found we end up with a set of
canonical variables 
$(q_i, p_i)$ and a canonical Hamiltonian $H$. The first class
constraints $G^a(p,q)$ obey the Poisson 
bracket algebra $\{ G^a, H \} = 0$, $\{G^a, G^b \} = f^{abc} G^c$,
where all structure functions $f^{abc}$ 
are regarded as independent of the canonical variables for
simplicity. To each first class constraint we associate a Lagrange
multiplier $\lambda^a$ and its canonical momentum $\Pi^a$. We also
introduce two sets of ghosts $(c^a, \overline{c}^a)$ and their
corresponding canonical momenta $(\overline{\cal P}^a, {\cal
  P}^a)$ satisfying $\{\overline{\cal P}^a, c^b \} = \{ {\cal P}^a,
\overline{c}^b \} = - \delta^{ab}$.  Then, the BRST charge can be
built as 
\be
\label{1}
Q = \int dx \,\, ( G^a c^a + \half f^{abc} \overline{\cal P}^a c^b c^c -
{\cal P}^a \Pi^a ),
\ee
where the integration is performed over the space coordinates. The
Hamiltonian BRST transformations are  
\bea
\label{2}
\delta c^a &=& - \half f^{abc} c^b c^c, \qquad \qquad \,\,\,\, \delta
\overline{c}^a = \Pi^a,  \\
\label{3}
\delta \overline{\cal P}^a &=& - G^a - f^{abc} \overline{\cal P}^b c^c,
\qquad \,\, \delta {\cal P}^a = 0, \\
\label{4}
\delta \lambda^a &=& {\cal P}^a, \qquad \qquad \qquad \qquad
\delta \Pi^a = 0,
\eea
and the BRST invariant action is 
\be
\label{5}
S = \int dx \,\, ( p_i \dot{q}_i + \dot{\cal P}^a \overline{c}^a +
\dot{c}^a \overline{\cal P}^a + \Pi^a \dot{\lambda}^a - H - \{Q,
\Psi\} ),
\ee
where a dot means time derivative and $\Psi$ is the gauge fixing
function. The main assertion in the BRST-BFV
formalism is that the path integral $Z= \int D[\phi] \exp(i S)$, where
$D[\phi]$ is the usual Liouville measure over all fields and ghosts,
is independent of the gauge fixing function $\Psi$.

Usually the gauge fixing function is chosen as 
\be
\label{6}
\Psi = \int dx \,\, \left[ \overline{c}^a \left( \half \xi \Pi^a +
    f^a(q) \right) + \overline{\cal P}^a \lambda^a \right],
\ee
where $f^a$ is a function of $q_i$ only and $\xi$ is a gauge fixing
parameter. This choice of $\Psi$  implements the gauge  
choice $f^a + \dot{\lambda}^a$ which leads to many well known gauges
like linear gauge conditions in Yang-Mills theory.  

Let us now perform the functional integration over the ghost
momenta. Integration over $\overline{\cal P}^a$ results in a delta
functional $\delta( {\cal P}^a  - D_0 c^a )$, where $D_0 c^a =
\dot{c}^a + f^{abc} \lambda^b c^c$. Integration over ${\cal P}^a$ then
sets ${\cal P}^a = D_0 c^a$. The resulting action is
\be
\label{7} 
S = \int dx \,\, \left[ p_i \dot{q}_i - H + \dot{\overline{c}}^a D_0 c^a +
\overline{c}^a \delta f^a + \lambda^a G^a - \Pi^a \left( \half \xi
  \Pi^a + f^a - \dot{\lambda}^a \right) \right],
\ee
which is invariant under the BRST transformations 
\bea
\label{8}
\delta c^a &=& - \half f^{abc} c^b c^c, \qquad \qquad \quad \,\, \delta
\overline{c}^a = \Pi^a,  \\
\label{9}
\delta \lambda^a &=& D_0 c^a, \qquad \qquad \qquad \qquad
\delta \Pi^a = 0.
\eea
Usually the integration over $p_i$ can be done if we identify the
Lagrange multiplier $\lambda^a$ with some variable in the original
Lagrangian formalism. For the Yang-Mills theory, for instance, the
Lagrange multiplier is just $A_0^a$ while $\Pi^a$ is the auxiliary
field needed to have off-shell nilpotency. Performing the integration
over $p_i$, the first two terms of the action (\ref{7}) together with
the term $\lambda^a G^a$ recompose the original Lagrangian and we are 
left with 
\be
\label{10}
S = \int dx \,\, \left[ L(q, \dot{q}) + \dot{\overline{c}}^a D_0 c^a +
\overline{c}^a \delta f^a - \Pi^a \left( \half \xi
  \Pi^a + f^a - \dot{\lambda}^a \right) \right].
\ee
This is the Lagrangian form of the gauge fixed action. It is invariant
under the BRST transformations (\ref{8}) and (\ref{9}). 

In \cite{Rivelles:1995gb} we particularised to the Yang-Mills case and
studied the symmetries of the gauge fixed action involving only the
ghost terms in (\ref{10}). We showed that in the new variables the BRST
transformations are nonlocal or even not manifestly covariant. Here we
want to focus in the sector $(\lambda^a, \Pi^a)$. It is easily seen
that the most general transformation leaving this sector of the action
invariant is 
\be
\label{11}
\Pi^{\prime a} = - \Pi^a - \frac{2}{\xi} (f^a - \dot{\lambda}^a ),
\qquad \qquad \lambda^{\prime a} = \lambda^a.
\ee
Notice that this transformation has Jacobian equal to minus one so it
does not affect the path integral 
measure. The action (\ref{10}) retains its form but the BRST
transformation for the new variable is now 
\be
\label{12}
\delta \Pi^{\prime a} = -\frac{2}{\xi} \left( \delta f^a - \partial_0
  D_0 c^a  \right).
\ee
The BRST transformation for $\overline{c}^a$ must also be expressed in
  terms of $\Pi^{\prime a}$. Doing that and dropping the prime in
  $\Pi^{\prime a}$ we have a new set of BRST transformations which
  leave the action (\ref{10}) invariant. It is given by
\bea
\label{13}
\delta c^a &=& - \half f^{abc} c^b c^c, \qquad \qquad \quad\,\,\, \delta
\overline{c}^a =  - \Pi^{a} - \frac{2}{\xi} (f^a - 
\dot{\lambda}^a ),  \\
\label{14}
\delta \lambda^a &=& D_0 c^a, \qquad \qquad \qquad \qquad
\delta \Pi^a = -\frac{2}{\xi} \left( \delta f^a - \partial_0
  D_0 c^a  \right).
\eea
These BRST transformations have also been found in
\cite{Lahiri:2001ti} \footnote{For comparison note that
  its gauge fixing condition $f^A$ corresponds to
  $f^a - \dot{\lambda}^a$ in this paper, while the auxiliary field
  $h^A$ corresponds to $\Pi^a$ and the ghost $\overline{\omega}^A$ is
  proportional $\overline{c}^a$. Our change of variables (\ref{11}) is
  the same transformation proposed there.}. We can understand
the origin of these transformations as new forms of the BRST symmetry.
They are nilpotent and on-shell they reduce properly to its
previous form (\ref{8}) and (\ref{9}). Either form of the
transformations is nilpotent showing that they realize the same BRST
symmetry. 

It is worth remarking that other symmetries, besides BRST symmetry,
can be found for gauge fixed actions. The best known example is the
vector supersymmetry which appears in the non-Abelian Chern-Simons
theory in Landau gauge \cite{Birmingham:1990ap}  which can be enlarged
to a contraction of the $D(2|1;\alpha)$ superalgebra
\cite{Damgaard:1990wm}. This symmetry gives rise to new Ward
identities which were used in \cite{Birmingham:1990ap} to relate the
gauge and ghost propagators. In the present case, the new form of the
BRST transformations can be used in conjunction which its usual
form. This may be 
useful from a technical point of view but the physical content is the
same as those found with the usual BRST transformations since there is
only one symmetry present. 

Although we have worked with an irreducible gauge theory and with
bosonic constraints, the formalism presented here can be easily 
generalized to the reducible case and to systems with bosonic and
fermionic constraints. 

This work was partially supported by Conselho Nacional de
Desenvolvimento Cient\'\i fico e Tecnol\'ogico (CNPq) and PRONEX under
contract CNPq 66.2002/1998-99.

\end{document}